\documentstyle[epsf]{article}
\newcounter{saveeqn}
\newcommand{\alpheqn}{\setcounter{saveeqn}{\value{equation}}
\stepcounter{saveeqn}\setcounter{equation}{0}
\renewcommand{\theequation}
      {\mbox{\arabic{saveeqn}\alph{equation}}}}
\newcommand{\reseteqn}{\setcounter{equation}{\value{saveeqn}}
\renewcommand{\theequation}{\arabic{equation}}}

\begin{document}

\title{Symplectic algorithm for constant-pressure molecular dynamics 
using a Nos\'e-Poincar\'e thermostat}

\author{Jess B. Sturgeon and Brian. B. Laird\cite{brian}\\
{\em Department of Chemistry and Kansas Institute for
         Theoretical and Computational Science }\\
   {\em  University of Kansas }\\
   {\em  Lawrence, Kansas 66045, USA}}
\date{\today}
\maketitle

\begin{abstract}
We present a new algorithm for isothermal-isobaric
molecular-dynamics simulation. The method uses an extended Hamiltonian
with an Andersen piston combined with the Nos\'e-Poincar\'e thermostat, 
recently developed by Bond, Leimkuhler and Laird [J. Comp. Phys., 
{\bf 151}, xxxx (1999)]. This Nos\'e-Poincar\'e-Andersen (NPA)
formulation has advantages over the Nos\'e-Hoover-Andersen approach 
in that the NPA is Hamiltonian and can take advantage of symplectic 
integration schemes, which lead to enhanced stability for long-time 
simulations. The equations of motion are integrated using a Generalized 
Leapfrog Algorithm and the method is easy to implement, symplectic, 
explicit and time reversible. To demonstrate the stability of the method 
we show results for test simulations using a model for aluminum. 
\end{abstract}

\section{Introduction}
Traditionally, molecular-dynamics simulations are performed using 
constant particle number $N$, volume $V$ and energy $E$.  However,
these are not usually the conditions under which experiments are 
done and there has been much attention to the development of
simulation methods designed to sample from other, experimentally more 
relevant  ensembles, such as constant temperature (canonical) and/or 
constant pressure\cite{Allen87,Frenkel97,Brown84}.  Some of the most 
popular and useful of these are those based on so-called ``extended'' 
Hamiltonians, i.e., Hamiltonians in which extra degrees of freedom 
have been added to the system in order to ensure that the trajectory 
samples from the statistical distribution corresponding to the desired 
thermodynamic conditions. 

For a constant pressure system, for example, Andersen\cite{Andersen80} 
introduced the volume $V$, along with its corresponding conjugate 
momentum $\pi_V$, as extra variables. The new variables are coupled to 
the system in such a way as to guarantee that the trajectory (if ergodic) 
samples from an isobaric statistical distribution. Similarly, to generate 
a constant temperature distribution, Nos\'e\cite{Nose84a} introduced a 
new mechanical variable $s$ (with conjugate momentum $\pi_s$) that couples 
into the system through the particle momenta and acts to effectively 
rescale time in such a way as to guarantee canonically distributed 
configurations. These two extensions can be combined to give a Hamiltonian 
whose trajectories can be shown to sample from an isothermal-isobaric 
ensemble.\cite{Nose84b}

This combined Nos\'e-Andersen (NA) Hamiltonian is given by
\begin{equation}
{\cal H}_{NA} = V^{-2/3} \sum \frac{p_{i}^{2}}{2 m_{i} s^{2}} 
+ U(V^{1/3} {\bf q}) + \frac{\pi_{V}^{2}}{2 Q_{V}} 
+ \frac{\pi_{s}^{2}}{2 Q_{s}} + g kT \ln{s} + P_{ext} V \;,
\label{hna}
\end{equation}

where $p_{i}$ is the conjugate momentum to the scaled position 
$q_{i} = V^{-1/3} r_{i}$, $P_{ext}$ is the external pressure and
 $g$ is given by $N_{f}+1$ where  $N_{f}$ is  number of degrees of 
freedom of the original system. The quantities $Q_V$ and $Q_s$
are the masses of the Andersen ``piston'' and the Nos\'e thermostat 
variable, respectively. 

The equations of motion for this system are
\alpheqn
\begin{eqnarray}
\dot{p_{i}}& =& -V^{1/3} \nabla_{i} U(V^{1/3} {\bf q}) \label{eoma}\\
\dot{q_{i}} & = & \frac{p_{i}}{s^{2} m_{i} V^{2/3}}\label{eomb} \\
\dot{\pi_{V}} & = & {\cal P} - P_{ext} \label{eomc}\\
\dot{V} &=& \pi_{v}/Q_{V} \label{eomd}\\
\dot{\pi_{s}}&= &V^{-2/3} s^{-3} \sum \frac{p_{i}^{2}}{m_{i}} - \frac{g kT}{s}\label{eome} \\
\dot{s}& =& \pi_{s}/Q_{s} \; ,\label{eomf}
\end{eqnarray}
\reseteqn
where the instantaneous pressure ${\cal P}$ is given by 
\begin{equation}
{\cal P} = \frac{2}{3V}\sum_i \frac{p^2_i}{2m_i V^{2/3} s^2} - 
\frac{1}{3V} \sum_i \frac{\partial U}{\partial q_i}  q_i 
\end{equation}

There are two major drawbacks to this approach: First, because of 
the time rescaling, the time variable in Nos\'e dynamics is not 
``real'' time, so any discretized trajectory generated by 
numerically integrating the Nos\'e equations of motion must be 
transformed back into ``real'' time, leading to the configurations 
that are spaced at unequal ``real''-time intervals. This is 
inconvenient for the construction of equilibrium averages, especially 
of dynamical quantities. Second, the Hamiltonian is not 
{\it separable}\cite{Sanz-Serna95} (that is, the kinetic and potential 
terms in the Hamiltonian are not functions only of momenta and 
position variables, respectively), making standard Verlet/leapfrog 
approaches inapplicable.

By a change of variables and a time rescaling of the equations of 
motion, Hoover\cite{Hoover85} derived new equations of motion that 
generate the same trajectories (for the exact solution) as the 
original Nos\'e Hamiltonian, but in real time. This Nos\'e-Hoover 
dynamics has become a standard method in molecular simulation.  However, 
the change of variables that links the Nos\'e Hamiltonian to the
Nos\'e-Hoover equations of motion is a non-canonical transformation - the 
total energy function of the system is still conserved, but it is no
longer a Hamiltonian, since the equations of motion cannot be derived
from it. Although a variety of very good time-reversible methods have
been put forward,\cite{Martyna94,Martyna96} the lack of Hamiltonian 
structure precludes the use of symplectic integration schemes, 
which have been shown to have superior stability over non-symplectic 
methods\cite{Sanz-Serna95}. 

\section{The Nos\'e-Poincar\'e-Andersen (NPA) Hamiltonian}

Recently, Bond, Leimkuhler and Laird\cite{Bond99} have developed a new
formulation of Nos\'e constant-temperature dynamics in which a Poincar\'e 
time transformation is applied directly to the Nos\'e Hamiltonian, 
instead of applying a time transformation to  the equations of motion 
as in Nos\'e-Hoover. The result of this is a method that runs in ``real'' 
time, but is also Hamiltonian in structure. In this work we  combine 
this new thermostat with the Andersen method for constant pressure to 
give an algorithm for isothermal-isobaric molecular dynamics.
For a system with an Andersen piston, the new Nos\'e-Poincar\'e-Andersen 
(NPA) Hamiltonian is given by
\begin{equation}
{\cal H}_{NPA} = [{\cal H}_{NA} - {\cal H}_{NA}(t=0)] s \; ,
                                                  \label{conserved}
\end{equation}
where ${\cal H}_{NA}$ is given in Eq.~\ref{hna}.
As discussed in Ref.~12, the above form of the Hamiltonian (a specific 
case of a Poincar\'e time transformation) will generate the same 
trajectories as the original Nos\'e-Andersen Hamiltonian, except with 
time rescaled by $s$ (which puts the trajectories back into real time), 
The resulting equations of motion (except for $\pi_{s}$) for this 
constant pressure and temperature Nos\'e-Poincar\'e Hamiltonian are 
the as those given above for the Nos\'e-Andersen system (Eqs.~\ref{eoma}-\ref{eomf}),
except that the right-hand side is multiplied by the thermostat 
variable $s$. For $\pi_{s}$ we have
\begin{equation}
\dot{\pi_{s}}= -s \frac{\partial {\cal H}}{\partial s} - \Delta H 
             =V^{-2/3} \sum \frac{p_{i}^{2}}{m_{i}s^{2}} - gkT - \Delta H
\end{equation}
where $\Delta {\cal H} \equiv {\cal H}_{NA} - {\cal H}_{NA}(t=0)$.

\section{Integrating the NPA Equations of motion}
The NPA Hamiltonian is nonseperable since the kinetic energy contains 
the extended ``position'' variables $s$ and $V$.  The equations of 
motion for a general time-independent, non-seperable Hamiltonian can  
be written (for general positions $Q$ and conjugate momenta $P$)
\begin{eqnarray}
\dot{Q} &= &G(P,Q) \nonumber \\
\dot{P} &= &F(P,Q)\; ,
                                                  \label{gla}
\end{eqnarray}
where $G(P,Q) = \frac{\partial {\cal H}}{\partial P}$ and 
$F(P,Q) = -\frac{\partial {\cal H}}{\partial Q}$. (For a seperable
Hamiltonian $G$ is only a function of $P$ and $F$ is only a function 
of $Q$.) For such a nonseperable system, standard symplectic 
splitting methods, such as the Verlet/leapfrog algorithm, are not
directly applicable. However, symplectic methods specifically for  
nonseperable systems have been developed\cite{Sanz-Serna95}.
One simple  example that is second-order and time-reversible is 
the Generalized Leapfrog Algorithm (GLA)
\begin{eqnarray} 
P_{n+1/2} & = & P_{n} + h F(P_{n+1/2},Q_{n})/2\nonumber \\
Q_{n+1} & = & Q_{n} + h[G(P_{n+1/2},Q_{n}) + G(P_{n+1/2},Q_{n+1})]/2\nonumber \\
P_{n+1} & = & P_{n+1/2} + h F(P_{n+1/2},Q_{n+1})/2 \; ,
\end{eqnarray}
where $h$ is the time step and $P_{n}$ and $Q_{n}$ are the 
approximations to $P(t)$ and $Q(t)$ at $t = t_{n} = nh$. (This 
method can be obtained as the concatenation of the Symplectic
Euler method, 
\begin{eqnarray} 
P_{n+1} & = & P_{n} + h F(P_{n+1},Q_{n})\nonumber \\
Q_{n+1} & = & Q_{n} + h G(P_{n+1},Q_{n})\;, 
\end{eqnarray}
with its {\it adjoint}\cite{Sanz-Serna95}.
The concatenation of a integrator with its adjoint guarantees a
time-reversible method.) This method is a simple example 
of a class of symplectic integrators for nonseperable 
Hamiltonians\cite{Sanz-Serna88,Lasagni88,Suris89,Sanz-Serna91}.

Applying the GLA to the NPA equations of motion gives
\alpheqn
\begin{eqnarray}
p_{i, n+1/2} &=& p_{i, n} - \frac{h}{2} s_{n}  V_{n}^{1/3} \, \nabla_i U(V^{1/3} {\bf q})  \\
\pi_{v, n+1/2} &=& \pi_{v, n} + \frac{h}{2} s_{n} [ {\cal P}({\bf q}_{n},{\bf p}_{n+1/2},V_{n},
s_{n}) - P_{ext} ] \\
\pi_{s, n+1/2} &=& \pi_{s, n} + \frac{h}{2} \left( \sum_{i=1}^{N} 
  \frac{p_{i, n+1/2}^{2}}{m_{i} V_{n}^{2/3} s^{2}_{n}} - g k_{B} T \right) \nonumber \\
  & & - \frac{h}{2} \Delta {\cal H}({\bf q}_{n},{\bf p}_{n+1/2},V_{n},\pi_{v,n+1/2},s_{n},\pi_{s,n+1/2}) 
                                                  \label{pishalf}\\
s_{n+1} &=& s_n + \frac{h}{2} (s_{n} + s_{n+1}) \frac{\pi_{s, n+1/2}}{Q_{s}} \\
V_{n+1} &=& V_{n} + \frac{h}{2} (s_{n} + s_{n+1}) \frac{\pi_{v, n+1/2}}{Q_{v}} \\
q_{i,n+1} &=& q_{i,n} + \frac{h}{2} \left( \frac{1}{s_{n}V_{n}^{2/3}} + 
  \frac{1}{s_{n+1}V_{n+1}^{2/3}} \right) \frac{p_{i, n+1/2}}{m_{i}} \\
\pi_{s, n+1} &=& \pi_{s, n+1/2} + \frac{h}{2} \left( \sum_{i=1}^{N} 
  \frac{p_{i, n+1/2}^{2}}{m_{i} V_{n+1}^{2/3} s^{2}_{n+1}} - g k_{B} T \right) \nonumber \\
  & & - \frac{h}{2} \Delta {\cal H}({\bf q}_{n+1},{\bf p}_{n+1/2},V_{n+1},\pi_{v,n+1/2},s_{n+1},\pi_{s,n+1/2})\\
\pi_{v, n+1} &=& \pi_{v, n+1/2} + \frac{h}{2} s_{n+1} [ {\cal P}({\bf q}_{n+1},{\bf p}_{n+1/2},V_{n+1},s_{n+1}) - P_{ext} ] \\
p_{i, n+1} &=& p_{i, n+1/2} + \frac{h}{2} s_{n+1} V_{n+1}^{1/3} \, 
\nabla_i U(V^{1/3}_{n+1}{\bf q_{n+1}})
\end{eqnarray}
\reseteqn

As in the case of the constant volume Nos\'e-Poincare algorithm, the 
GLA for the NPA is explicit - this is not necessarily the case for a  
general nonseperable Hamiltonian. Note that Eq.~\ref{pishalf} requires 
the solution of a scalar quadratic equation for $\pi_{s,n+1/2}$. Details 
of how to solve this equation without involving subtractive cancellation 
can be found in Ref.~12.

\section{Test Simulation Results and Summary}

In order to evaluate this method, simulations were performed using an 
embedded atom potential for aluminum\cite{Mei92}.  We report test 
simulations on a system of 256 particles with periodic boundary conditions 
for an aluminum melt at $T = 1000K$ and $P = 0$.
For this model  mass is measured in amu, distance in \AA \, and energy in 
eV, the natural time unit of the simulation is then 10.181fs, that is, a 
simulation time step of 0.1 corresponds to an acutal time step of 1.0181fs.  

\begin{figure}[!h]
   \centering
   \epsfxsize = 5.5in
   \epsfbox{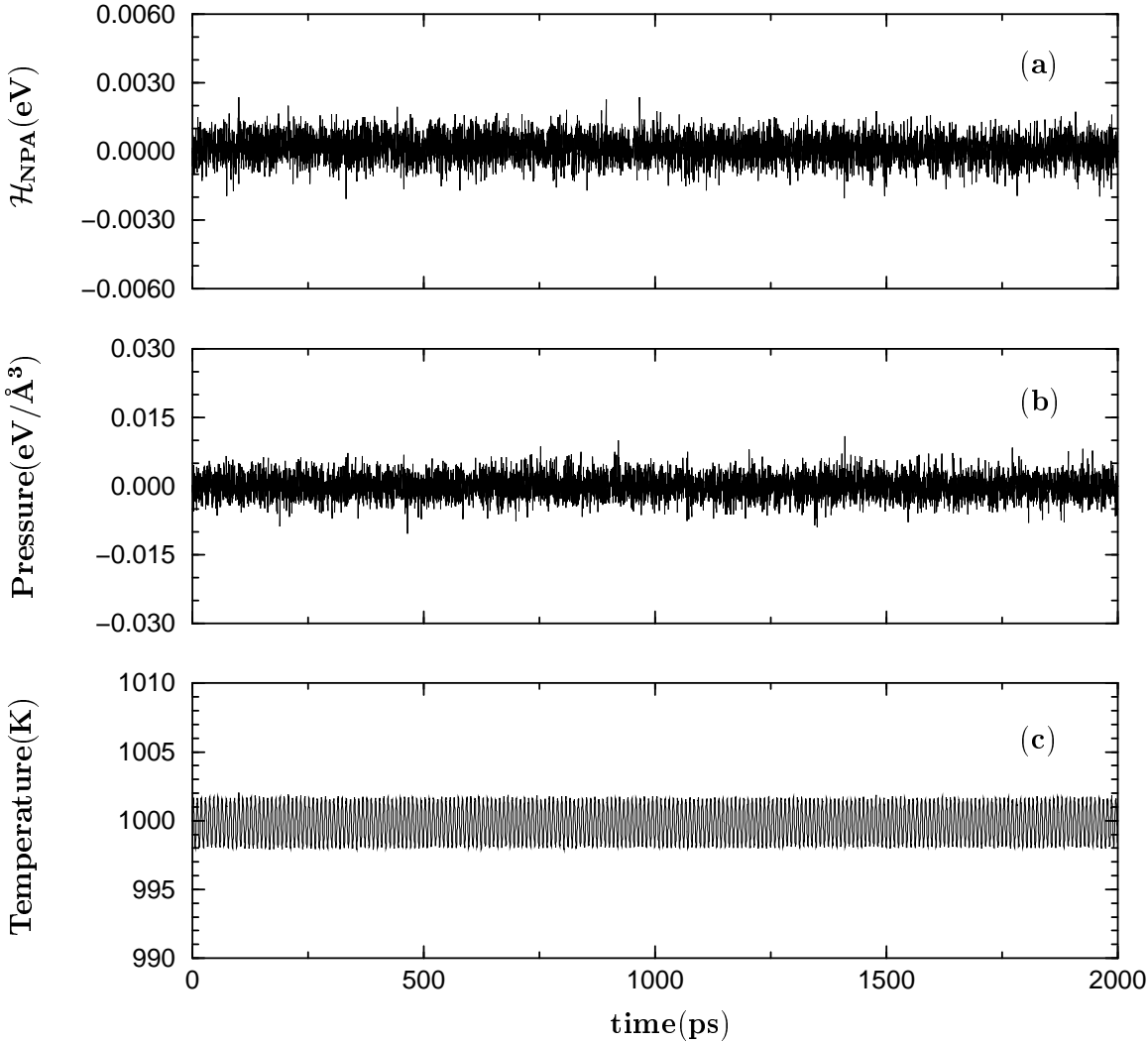}

\caption{\footnotesize (a) The value of the NPA Hamiltonian as a function 
of time for a long run (over 2ns) on a 256 particle aluminum system. The 
starting configuration had been equilibrated at $T=1000$K and $P=0$. The 
values of $Q_v$ and $Q_s$ (in reduced units) are 0.4 and $10^{-4}$, 
respectively. The pressure and temperature trajectories for this run are 
given in (b) and (c), respectively.} 
                                                  \label{fig:consqua}
\end{figure}

First the stability of the method was tested.  Fig.~\ref{fig:consqua}(a) 
shows the value of the NPA Hamiltonian (a conserved quantity) as a function 
of time in a long run.  The trajectory shown here was begun after initial 
equilibration at 1000K for 2$\times 10^6$ time steps (2.03ns). In this
simulation, the values of $Q_v$ and $Q_s$ (in reduced units) were 10$^{-4}$
and 2.5, respectively. The stability of the method is excellent, giving
no noticeable drift in ${\cal H}_{\mbox {\small NPA}}$ over the course of 
a long trajectory. The pressure and temperature trajectories for this run 
are also shown in Figs.~\ref{fig:consqua}(b) and~\ref{fig:consqua}(c), 
respectively.

\vspace{1cm}

In Fig. 2(a)-(c), we show the ability of the method to regulate the 
pressure, temperature and density, respectively, for three sets of extended 
variable masses. (The values  for $Q_s$ were larger here than that used in 
Figure~\ref{fig:consqua} because small values of that variable lead to 
instabilities when the initial temperature is far from the target 
temperature.) The system was initialized to an fcc lattice of initial 
density 0.06021\AA$^{-3}$ with the individual velocity  components chosen 
from a Maxwell-Boltzmann  distribution at 100K. The simulation was then run 
with the NPA with $T=1000K$ and $P=0$.  In all cases, the instantaneous 
pressure, temperature and density evolve quickly and stabilize about their 
desired values.

\vspace{1cm}

\begin{figure}[!h]
   \centering
   \epsfxsize = 5.5in
   \epsfbox{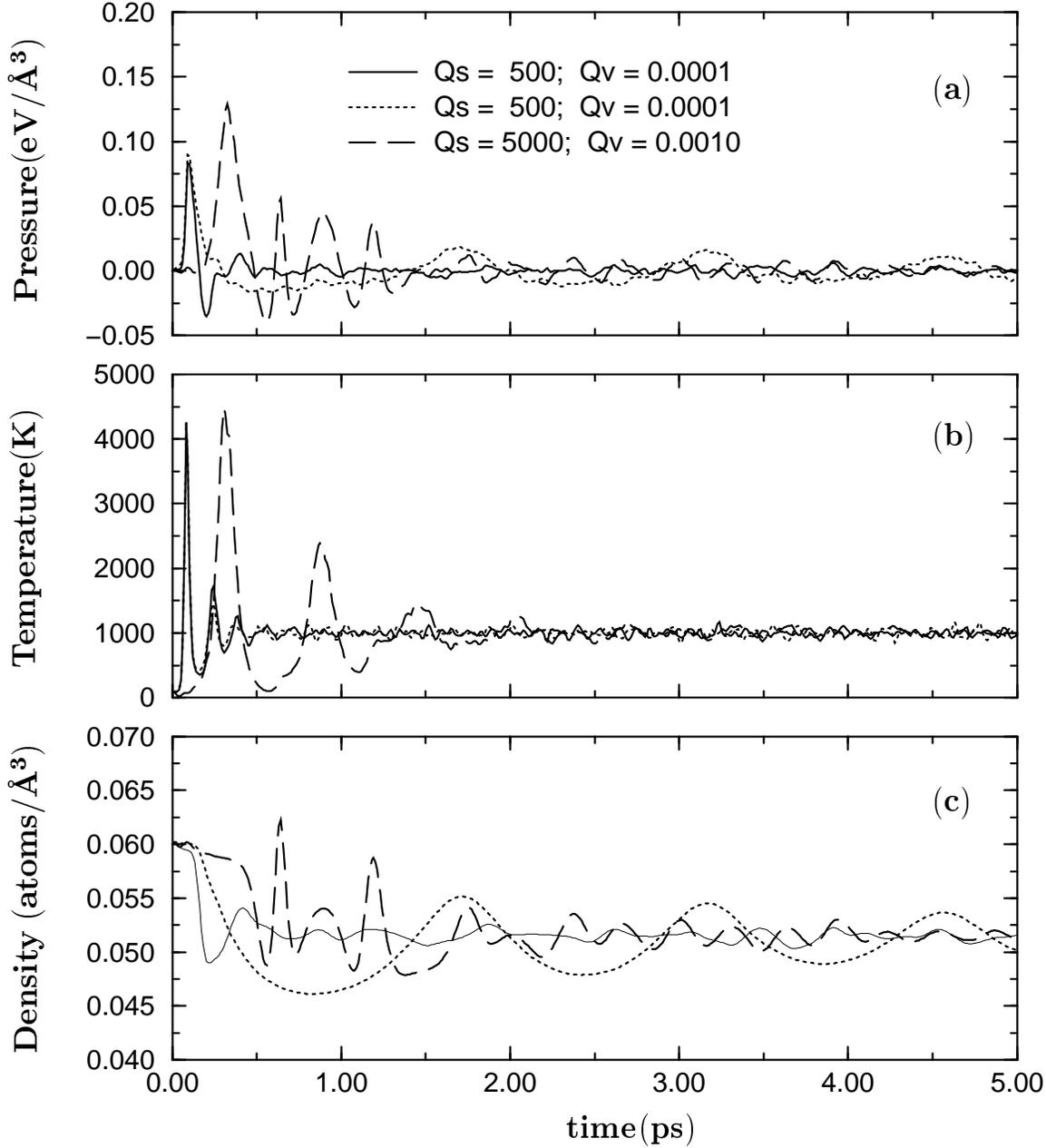}

\caption{\small Pressure (a), temperature (b) and density (c)trajectories 
  for the 256 particle aluminum system using the NPA algorithm with $T=1000K$ 
  and $P=0$ starting from an initial configuration in an fcc lattice with
  density $\rho = 0.06021\AA^{-3}$ and initial velocities chosen from a 
  Maxwell-Boltzmann distribution at 100K.}
                                                  \label{fig:fig2}
\end{figure}

\begin{figure}[!h]
   \epsfxsize = 2.2in
   \epsfbox{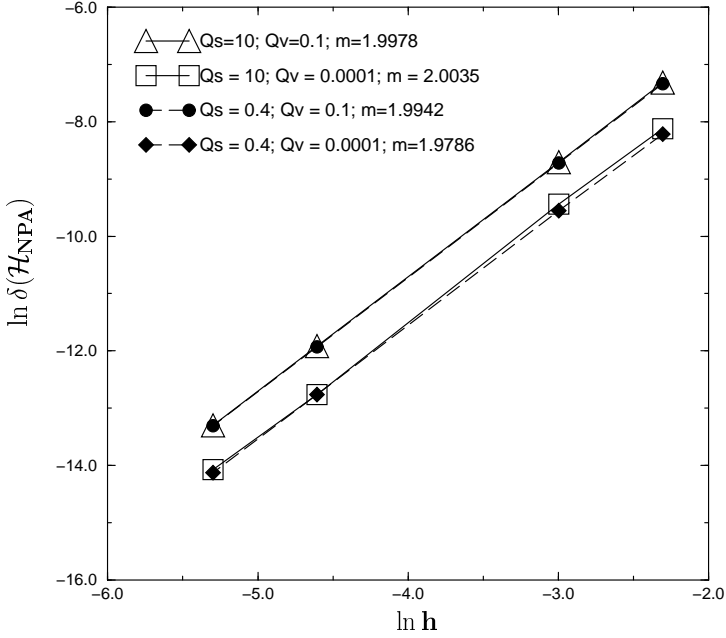}

\caption{\small Log-log plot of the energy error $\delta{\cal H}_{NPA}$  
  versus the time step for a variety of piston and thermostat masses.  
  The order of the error is given by the slope of lines and is labeled 
  as $m$ in the legend.}
                                                  \label{fig:sigmadt}
\end{figure}

The GLA has a global error that is second-order in the time step.  To
demonstrate that this also is true in our results, a series of simulations 
were performed using various values for the  time step. The system was 
initialized in an identical manner to that described in the last paragraph 
and then run for a total time of 2.0362 ps.  Figure [\ref{fig:sigmadt}] 
shows, for several combinations of extended variable masses, a log-log plot
of the energy error, as estimated by the standard deviation of 
$\cal{H}_{\mbox {\small NPA}}$ (Eq.~\ref{fig:sigmadt}), versus the time step. 
One notes that here smaller values of $Q_v$ lead to smaller fluctuations in 
${\cal H}_{\mbox {\small NPA}}$.

\vspace{4cm}

\vspace{4cm}
\begin{figure}[!h]
   \epsfxsize = 2.2in
   \epsfbox{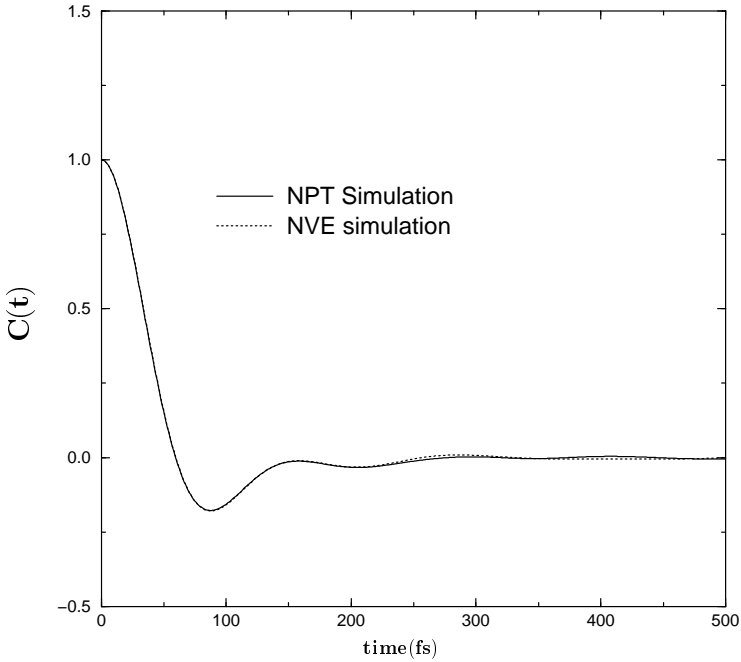}

\caption{\small  The normalized velocity autocorrelation function, $C(t)$, 
  for the embedded-atom model for aluminum  calculated for a 256 particle 
  system at T=1000K and P=0. The solid line is calculated using the NPA 
  algorithm described herein. The dotted line is a constant NVE simulation 
  under similar conditions.}
                                                  \label{fig:vacf}
\end{figure}

Finally, to demonstrate that the method yields relevant dynamical 
quantities, the normalized velocity autocorrelation function, 
$C(t) = \langle{\bf v}(t)\cdot{\bf v}(0)\rangle/\langle{\bf v}(0)\cdot{\bf v}(0)\rangle$, 
was calculated using our constant NPT algorithm (with $Q_v$ and $Q_s$ as in
Fig. 1) and compared to the same quantity calculated using standard constant 
NVE molecular dynamics (with a velocity-Verlet integrator\cite{Swope82}). 
The NVE simulations were run at an energy and density corresponding to 
the average energy and density for the constant NPT simulations.  This 
comparison is shown in Figure [\ref{fig:vacf}].  Both systems were first 
equilibrated at 1000K for 200,000 steps (203.6 ps) and run for 20,000 steps 
(20.36 ps) to collect averages. $C(t)$ for the Nos\'e-Poincar\'e-Andersen  
method for constant NPT molecular dynamics is indistinguishable in this
figure from that of the NVE simulation.

\section{Acknowledgements}
We gratefully acknowledge Steve Bond and Ben Leimkuhler for helpful 
conversations and invaluable advice, as well as the Kansas Center for 
Advanced Scientific Computing for the use of their computer facilities. 
We also would like to thank the National Science Foundation (under grants
CHE-9500211 and DMS-9627330) as well as the University of Kansas
General Research Fund for their generous support of this research.


\begin{thebibliography}{10}
\bibitem{brian}[*] Author to whom correspondence should be addressed.

\bibitem{Allen87}
M.A. Allen and D.J. Tildesley, {\em Computer Simulation of Liquids}, (Oxford
  Science Press, Oxford, 1987).

\bibitem{Frenkel97}
D.~Frenkel and B.~Smit, {\em Understanding Molecular Simulation}, (Academic
  Press, New York, 1996).

\bibitem{Brown84}
D.~Brown and J.~H.~R. Clarke, Mol. Phys. {\bf 51}, 1243--1252 (1984).

\bibitem{Andersen80}
H.C. Andersen, J. Chem. Phys {\bf 72}, 2384--2393 (1980).

\bibitem{Nose84a}
S.~Nose, Mol. Phys. {\bf 52}, 255 (1984).

\bibitem{Nose84b}
S.~Nose, J. Chem. Phys. {\bf 81}, 511 (1984).

\bibitem{Sanz-Serna95}
J.M. Sanz-Serna and M.P Calvo, {\em Numerical Hamiltonian Problems}, (Chapman
  and Hall, New York, 1995).

\bibitem{Hoover85}
W.G. Hoover, Phys. Rev. A {\bf 31}, 1695 (1985).

\bibitem{Martyna94}
D.J.~Tobias G.J.~Martyna and M.L.~Klein, J. Chem. Phys. {\bf
  101}, 4177--4189 (1994).

\bibitem{Martyna96}
D.J.~Tobias G.J.~Martyna, M.E.~Tuckerman and Michael~L. Klein,
  Mol. Phys. {\bf 87}, 1117--1157 (1996).

\bibitem{Bond99}
S.D.~Bond, B.J.~Leimkuhler, and B.B.~Laird, J. Comp. Phys.
  {\bf 151}, xxxx (1999).

\bibitem{Sanz-Serna88}
J.M.~Sanz-Serna, BIT {\bf 28}, 877--883 (1991).

\bibitem{Lasagni88}
F.~Lasagni, ZAMP {\bf 39}, 952--953 (1988).

\bibitem{Suris89}
Y.B. Suris, U.S.S.R. Comput. Maths. Math. Phys. {\bf 29}, 138--144 (1989).

\bibitem{Sanz-Serna91}
J.M. Sanz-Serna, Acta Numer. {\bf 1}, 243--286 (1991).

\bibitem{Mei92}
J.~Mei and J.W. Davenport, Phys. Rev. B {\bf 46}, 21--27 (1992).

\bibitem{Swope82}
W.C.~Swope, H.C.~Andersen, P.H.~Berens, and K.R.~Wilson, J. Chem. Phys. {\bf
  76}, 637 (1982).

\end{thebibliography}

\end{document}